\input epsf
\def\P{I\!\!P}

\magnification\magstephalf
\overfullrule 0pt
\def\gsim{\raise.3ex\hbox{$\;>$\kern-.75em\lower1ex\hbox{$\sim$}$\;$}}

\font\rfont=cmr10 at 10 true pt
\font\ritfont=cmti10 at 10 true pt
\def\ref#1{$^{\hbox{\rfont {[#1]}}}$}


\font\fourteenbf=cmbx12 scaled\magstep2

\font\tenbfit=cmbxti10
\font\sevenbfit=cmbxti10 at 7pt
\font\fivebfit=cmbxti10 at 5pt
\newfam\bfitfam 
\textfont\bfitfam=\tenbfit  \scriptfont\bfitfam=\sevenbfit
\scriptscriptfont\bfitfam=\fivebfit

\font\tenbfit=cmbxti10
\font\sevenbfit=cmbxti10 at 7pt
\font\fivebfit=cmbxti10 at 5pt
\newfam\bfitfam 
\textfont\bfitfam=\tenbfit  \scriptfont\bfitfam=\sevenbfit
\scriptscriptfont\bfitfam=\fivebfit

\font\tenbit=cmmib10
\newfam\bitfam
\textfont\bitfam=\tenbit%

\font\tenmbf=cmbx10
\font\sevenmbf=cmbx7
\font\fivembf=cmbx5
\newfam\mbffam
\textfont\mbffam=\tenmbf \scriptfont\mbffam=\sevenmbf
\scriptscriptfont\mbffam=\fivembf

\font\tenbsy=cmbsy10
\newfam\bsyfam 
\textfont\bsyfam=\tenbsy%


\def\pmb#1{\setbox0=\hbox{#1}
 \kern.05em\copy0\kern-\wd0 \kern-.025em\raise.0433em\box0 }

\def\slash{/\kern-.5em}

\def \half {{\textstyle {1 \over 2}}}

 %


\def\boxit#1{\vbox{\hrule\hbox{\vrule\kern1pt\vbox
{\kern1pt#1\kern1pt}\kern1pt\vrule}\hrule}}

\def\h{\hfill\break}
\parskip=6pt
\parindent=0pt
\hsize=17truecm\hoffset=-5truemm
\vsize=23truecm
\def\footnoterule{\kern-3pt
\hrule width 17truecm \kern 2.6pt}


\catcode`\@=11 

\def\nolabels{\def\wrlabeL##1{}\def\eqlabeL##1{}\def\reflabeL##1{}}
\def\writelabels{\def\wrlabeL##1{\leavevmode\vadjust{\rlap{\smash%
{\line{{\escapechar=` \hfill\rlap{\sevenrm\hskip.03in\string##1}}}}}}}%
\def\eqlabeL##1{{\escapechar-1\rlap{\sevenrm\hskip.05in\string##1}}}%
\def\reflabeL##1{\noexpand\llap{\noexpand\sevenrm\string\string\string##1}}}
\nolabels
\global\newcount\refno \global\refno=1
\newwrite\rfile
\def\defref{$^{{\hbox{\rfont [\the\refno]}}}$\nref}
\def\nref#1{\xdef#1{\the\refno}\writedef{#1\leftbracket#1}%
\ifnum\refno=1\immediate\openout\rfile=refs.tmp\fi
\global\advance\refno by1\chardef\wfile=\rfile\immediate
\write\rfile{\noexpand\item{#1\ }\reflabeL{#1\hskip.31in}\pctsign}\findarg}
\def\findarg#1#{\begingroup\obeylines\newlinechar=`\^^M\pass@rg}
{\obeylines\gdef\pass@rg#1{\writ@line\relax #1^^M\hbox{}^^M}%
\gdef\writ@line#1^^M{\expandafter\toks0\expandafter{\striprel@x #1}%
\edef\next{\the\toks0}\ifx\next\em@rk\let\next=\endgroup\else\ifx\next\empty%
\else\immediate\write\wfile{\the\toks0}\fi\let\next=\writ@line\fi\next\relax}}
\def\striprel@x#1{} \def\em@rk{\hbox{}} 
\def\lref{\begingroup\obeylines\lr@f}
\def\lr@f#1#2{\gdef#1{\defref#1{#2}}\endgroup\unskip}
\newwrite\lfile
{\escapechar-1\xdef\pctsign{\string\%}\xdef\leftbracket{\string\{}
\xdef\rightbracket{\string\}}}

\def\writestop{\def\writestoppt{\immediate\write\lfile{\string\p
ageno%
\the\pageno\string\startrefs\leftbracket\the\refno\rightbracket%
\string\def\string\secsym\leftbracket\secsym\rightbracket%
\string\secno\the\secno\string\meqno\the\meqno}\immediate\closeout\lfile}}
\def\writestoppt{}\def\writedef#1{}
\catcode`\@=12 
\centerline{\fourteenbf Soft diffraction dissociation}
\vskip 8pt
\centerline{A Donnachie}
\centerline{Department of Physics, Manchester University$^*$}
\vskip 5pt
\centerline{P V Landshoff}
\centerline{DAMTP, Cambridge University$^*$}
\footnote{}{$^*$ email addresses: ad@a35.ph.man.ac.uk, \ pvl@damtp.cam.ac.uk}
\bigskip
{\bf Abstract}

We refute past suggestions that the data on soft diffraction dissociation
demand modifications to conventional Regge theory.

\vskip 15truemm

Hard diffractive processes have been an active area of study at HERA
and the Tevatron, and will continue to be so at the CERN LHC, but one
must be uneasy about all theoretical analyses if we do not even understand
soft diffraction.
The data\defref\albrow{
M G  Albrow et al, Nuclear Phys.B108 (1976) 1
}\defref\arm
{J C M  Armitage et al, Nuclear Physics B194 (1982) 365
}
from the CERN ISR for soft diffraction dissociation,
$$
pp\to pX
\eqno(1)
$$ 
with the proton emerging with almost no energy loss, 
have been shown\defref\robroy{
D P Roy and R G Roberts, Nuclear Physics B77 (1974) 240
}\defref\diffdis{
A Donnachie and P V Landshoff, Nuclear Physics  B244 (1984) 322
}
to be in good agreement with the triple-Regge description\defref\goul{
K Goulianos, Physics Letters B358 (1995) 379
}\defref\book{
A Donnachie, H G Dosch, P V Landshoff and O Nachtmann, {\it Pomeron physics
and QCD}, Cambridge University Press (2002). See 
www.damtp.cam.ac.uk/user/pvl/QCD
}. However, the higher-energy data from the CERN SPS collider\defref\uaf{
UA4 collaboration, M Bozzo et al, Physics Letters B136 (1984) 217
}\defref\uae{
UA8 collaboration, A Brandt et al, Nuclear Physics B514 (1998) 3
}
do not show the increase of the cross section with energy expected from
simple fits\ref{\book}.
This has led certain authors\ref{\goul}\defref\es{
S Erhan and P E Schlein, Physics Letters B481 (2000) 177
}
to introduce unconventional features into the data analysis, which
yield energy dependence in conflict with the standard notions
of Regge theory\defref\collins
{P D B Collins, {\it Introduction to Regge Theory and High Energy Physics},
Cambridge University Press (1977)
}\ref{\book}.
In this paper, we show that this is unnecessary and
that conventional Regge theory gives an adequate description of the data.

Let $\xi =1-x_F$ be the fractional energy loss of the emerging fast proton (or
antiproton) and $t$ the squared momentum transfer to it. 
According to standard Regge theory,
when $\xi$ is small enough and the energy large enough
the reaction (1) is generated through the
very fast proton ``radiating'' a pomeron, which hits the other proton and
breaks it up. This gives
$$
{d^2\sigma \over dt\, d\xi}=D(t, \xi)~\sigma ^{\P p}(M^2,t)
\eqno(2)
$$
where $M$ is the invariant mass of the system $X$ in (1),
$$
M^2\sim\xi s
\eqno(3)
$$
This factorisation of the differential cross section into a product of a
``pomeron flux'' factor $D(t, \xi)$ and a pomeron-proton ``cross section''
$\sigma ^{\P p}$ is ambiguous; different definitions in the 
literature\ref{\diffdis}\defref\berger{
E L Berger, J C Collins, D Soper and G Sterman, Nuclear Physics B288 (1987) 704}
differ by a factor of ${1\over 2}\pi$ in the definition of $D(t, \xi)$, 
with a corresponding difference in the definition of $\sigma ^{\P p}$.
Our own definition is\ref{\diffdis}
$$
D(t,\xi)={9\beta_{\P}^2\over 4\pi^2}\,(F_1(t))^2\,\xi^{1-2\alpha_{\P}(t)}
\eqno(4)
$$
where $F_1(t)$ is the proton's Dirac form factor,
$$
\alpha_{\P}(t)=1+\epsilon_{\P}+\alpha'_{\P}t
$$$$
\epsilon_{\P}=0.0808~~~~~~~~~\alpha'_{\P}= 0.25 \hbox{ GeV}^{-2}
\eqno(5)
$$
is the pomeron trajectory  and 
$\beta_{\P}^2\approx 3.5$ GeV$^{-2}$ is defined such that the pomeron-exchange
contribution to the total cross section is\defref\sigtot{
A Donnachie and P V Landshoff, Physics Letters B296 (1992) 227
}
$$
\sigma^{pp}(s)\sim 18\, \beta_{\P}^2 (\alpha'_{\P}s)^{\epsilon_{\P}}
\cos(\half\pi\epsilon_{\P})=21.7s^{0.0808}
\eqno(6)
$$
in mb-GeV units.

Experimentalists commonly integrate
${d^2\sigma / dt\, d\xi}=D(t, \xi)~\sigma ^{\P p}(M^2,t)$
over all $t$ and down to some value $M_{\hbox{\fiverm{MIN}}}$ of $M$, and refer to
a total diffractive cross section 
$$
\sigma^{\hbox{{\fiverm DIFF}}}(s;M_{\hbox{\fiverm{MIN}}},\xi_{\hbox{\fiverm{MAX}}})=
\int _{M=M_{\hbox{\fiverm{MIN}}}}^{\xi=\xi_{\hbox{\fiverm{MAX}}}}dt\, d\xi
\,d^2\sigma / dt\, d\xi
\eqno(7)
$$ 
The contribution (2) makes 
$\sigma^{\hbox{{\fiverm DIFF}}}(s;M_{\hbox{\fiverm{MIN}}})$  rise
with $s$ almost as fast as $s^{2\epsilon_{\P}}$. As this is faster than
the total cross section, 
it has been claimed\defref\gotsman{
E Gotsman, E M Levin and U Maor, Physical Review D49 (1994) 4321
}
that shadowing effects are important. However,
at present energies 
$\sigma^{\hbox{{\fiverm DIFF}}}(s;M_{\hbox{\fiverm{MIN}}})$ is much 
less than the total cross section, and so
it seems to us likely that shadowing is not yet an issue. This
is controversial: for a number of reasons we believe\ref{\book} that
shadowing is still unimportant also in forward elastic scattering
and in the total cross section, but this view is not universal.

Being a hadronic cross section, $\sigma ^{\P p}(M^2,t)$ is expected to
be dominated by pomeron exchange when the corresponding centre-of-mass
energy $M$ is sufficiently large, and then it becomes proportional to
$(M^2)^{\epsilon_{\P}}$, in analogy with the behavior $s^{\epsilon_{\P}}$
of $\sigma^{pp}(s)$ given in (6).
When the centre-of-mass energy $\sqrt s$ for $\sigma^{pp}(s)$ or $M$ 
for $\sigma ^{\P p}(M^2,t)$ is not large enough, pomeron
exchange alone is not sufficient, and nonleading Regge exchanges become
important. The dominant such exchange in $\sigma^{pp}(s)$ is $f_2$ 
exchange\ref{\book}, though $\rho,\omega$ and $a_2$ exchanges are present 
too. Each of these contributes to $\sigma^{pp}(s)$ a term behaving
approximately as $1/\sqrt s$. When this is added to the slowly-rising
pomeron-exchange term (6), the result is that, as $s$ increases,
$\sigma^{pp}(s)$ initially falls with increasing $s$, and then begins to 
rise. We expect that $f_2$  exchange will similarly contribute
to $\sigma ^{\P p}(M^2,t)$, so that it too falls initially with increasing
$M$, and then begins to rise. Note also the very important point that,
for $s$ less than some value $s_0$, these exchanges are not sufficient.
For most
hadron-hadron total cross sections the value of $\sqrt s_0$ 
is found\defref\cudell{
J R Cudell et al, Physical Review D61 (2000) 034019; {\it erratum} D63 (2001) 059901
}
to be about 4 to 6 GeV. Similarly, we should expect the simple exchange
theory to break down for $\sigma ^{\P p}(M^2,t)$ when $M$ is less
than some value $M_0$. One cannot know what the value of $M_0$ is;
we will be fairly bold and assume a value 3 GeV.
This limitation is important when one is interested in the 
total diffractive cross section $\sigma^{\hbox{{\fiverm DIFF}}}(s;M_{\hbox{\fiverm{MIN}}})$.
Usually the value of $M_{\hbox{\fiverm{MIN}}}$ that is chosen is less than the likely
value of $M_0$, in which case the integration over $M$ extends into a
region where there is poor theoretical understanding. 

\topinsert
\centerline{\epsfxsize=40truemm\epsfbox[0 0 270 200]{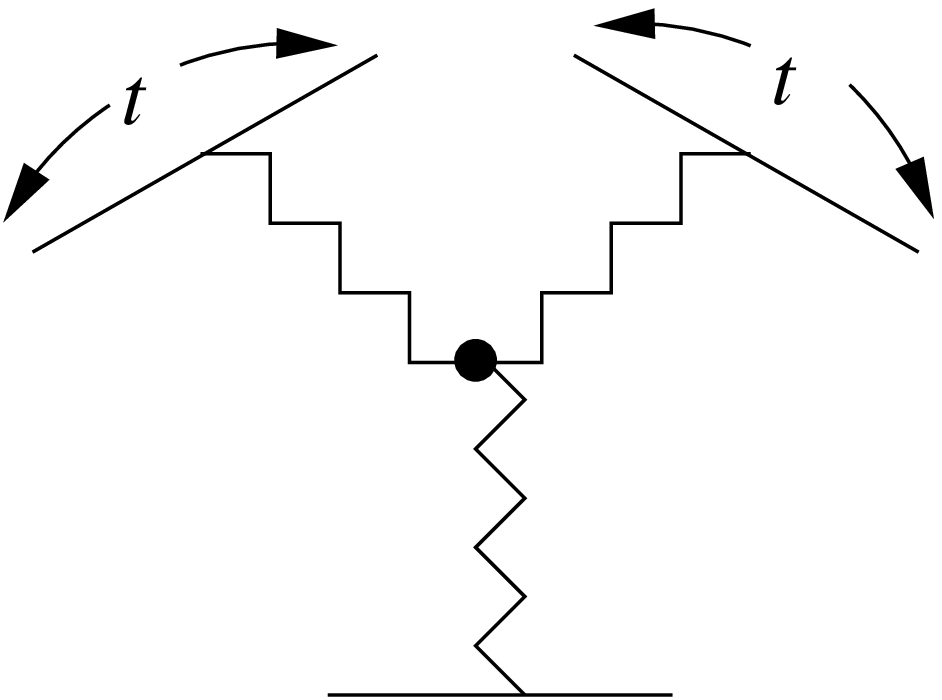}}\h
\centerline{\rfont Figure 1: triple-reggeon diagram}
\endinsert

According to what we have said,
when $M$ is large enough for $\sigma ^{\P p}(M^2,t)$ to be approximated
by pomeron and other reggeon exchanges, ${d^2\sigma / dt\, d\xi}$ in
(2) is the sum of triple-reggeon contributions. See figure 1. In this
figure, the two upper zigzag lines are the pomeron ``radiated'' by the
very-fast proton. The lower one is the pomeron or
reggeon exchanged in $\sigma ^{\P p}(M^2,t)$. When $\xi$ is not very small,
there are other contributions from a reggeon other than the pomeron
being radiated by the very-fast proton. Also, these additional contributions
can interfere with the pomeron-radiation term. So we obtain a sum of
terms of the type shown in figure 1:
$$
{\P\P\atop\P}~~~~~~{\P\P\atop f_2}
~~~~~~{f_2f_2\atop\P}~~~~~~{f_2f_2\atop f_2}
~~~~~~{f_2\P\atop\P}~~~~~~{\P f_2\atop\P}
~~~~~~{f_2\P\atop f_2}~~~~~~
{\omega\P\atop\omega}~~~~~\dots
\eqno(8)
$$
The last four terms are examples of interference terms.
All the terms but the last contribute equally to
$pp\to pX$ and $\bar pp\to \bar pX$ or $p\bar p\to pX$. However the present 
data turn out not to be sensitive to these terms so we treat $pp\to pX$ and 
$\bar pp\to \bar pX$ equally. There are also contributions from $a_2$
and $\rho$ exchange though, as we have said, these couple more weakly to 
the proton or antiproton. For simplicity we assume that the reggeon exchanges
are degenerate. At small $t$, it is important also to include
$$
{\pi\pi\atop\P}~~~~~~{\pi\pi\atop f_2}
\eqno(9)
$$

Altogether, there are a very large number of terms. For each there is
a triple-reggeon vertex, the blob in the centre of figure 1, whose 
dependence on $t$ is completely unknown. With so much freedom, it is
surely possible to obtain a good description of the data without having to
introduce new and unconventional devices into the fit.

There is a huge amount of data on soft diffraction dissociation.
If one 
plots them one finds that a lot of them are less than good. ISR and SPS
collider data have resolution problems for small $\xi$ and so we have
restricted our use of these data to $\xi\ge 0.02$. It is a great pity that
some of the SPS collider data for 
${d^2\sigma / dt\, d\xi}$
survives only in integrated form. It is unfortunate that none has been 
published from the Tevatron,
though Goulianos and Montanha\defref\goultwo{
K Goulianos and J Montanha, Physical Review D59 (1999) 114017
}
have reconstructed CDF data at $-t=0.05$ GeV$^2$ using a particular model.
There are some apparently very accurate fixed-target data\defref\sch{
R D Schamberger et al, Physical Review D17 (1978) 1268
}
though this data set has certain internal peculiarities
and at some values of $t$ its  normalisation disagrees with that of references
[{\albrow}] and [{\arm}], which agree with each other. Figure 2 shows the
fixed-target data at two values of $s$ and illustrates that any successful fit
must have quite a complicated structure. The differential 
cross section at fixed $\xi$ decreases with increasing $s$ in this energy range,
which is the opposite behaviour from that of the term ${\P\P\atop\P}$ in (8).
At very small $t$ the differential cross section
is very steep  and the data cannot be parametrised with a single
simple exponential $e^{bt}$, although this fits satisfactorily at larger
$t$. Simple exponential fits are often used by experimentalists to extrapolate
their data beyond the region where they are measured, but this is known to
be risky. For example, it is known\defref\esevten{
E710 collaboration: N A Amos et al, Physics Letters B186 (1987) 227
}
that the slope $b$ decreases rapidly as $\xi$ increases.

We have chosen to use collider data from reference
[{\arm}] at $s=2880$ GeV$^2$. They are relatively smooth, have quite small
error bars, and extend over a wide range of $t$, from $-t=0.11$ 
to $0.84$ GeV$^2$. Note, though, that it
may be unsafe
to extend the simple Regge-exchange picture to such a large value of $t$.
As we said,  to avoid resolution problems we restrict
ourselves to $\xi\ge 0.02$ with these data, and at this high energy
this corresponds to $M > 7.6$ GeV, comfortably satisfying our
condition $M>M_0$ when $M_0=3$ GeV.
To give leverage in $s$, we use SPS collider data from reference [{\uaf}]
at $-t=0.55$ and 0.75 GeV$^2$. 
We use also all the fixed-target data from reference [{\sch}];
they are at $s=262, 309, 366, 565, 741$
GeV$^2$ and
small $t$, $-t=0.036, 0.075, 0.131$ and $0.197$ GeV$^2$.
For these low-energy data the constraint $M>M_0$ is important. We use also
the reconstructed CDF data\ref{\goultwo} at $-t=0.05$ GeV$^2$. 

A simple least-$\chi^2$ fit to these data is not adequate because
the error bars on the ISR data are so much larger than on the fixed-target
data. So we artificially reduce the error bars on the smallest-$t$ ISR data
when we make the fit. Also, if we use only the reconstructed CDF data
at the one value of $t$ for which it is available, when we integrate the
fit over $\xi$ its $t$ dependence is in conflict with what we believe
to have been measured by CDF. So we have further reconstructed some
CDF data for ourselves, at $-t=0.01$ and 0.09 GeV$^2$ with
$\sqrt s=1800$ GeV.
\topinsert
\centerline{\epsfxsize=0.4\hsize\epsfbox[65 60 390 285]{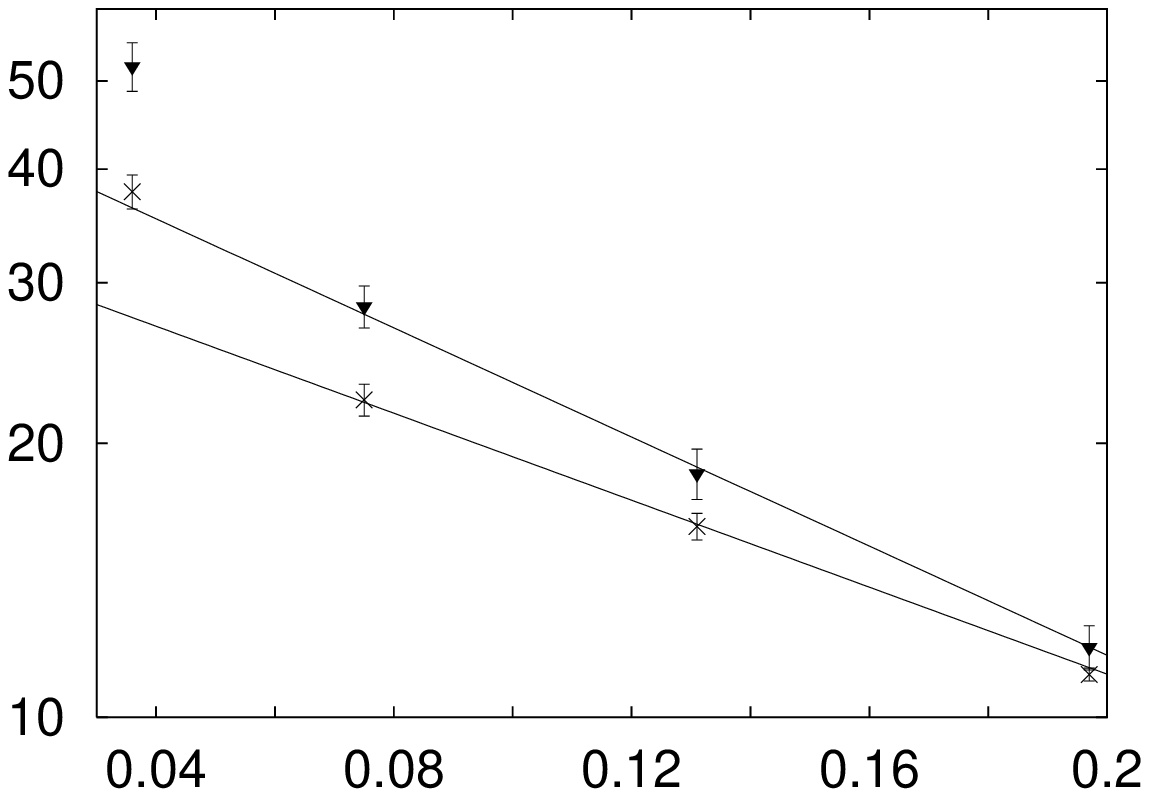}}
\vskip 2truemm
{\rfont Figure 2: data for ${(1/\pi)d^2\sigma / dt\, d\xi}$ in
mb~GeV$^{-2}$ from reference~%
[{\sch}] at $\xi=0.04$ for $s=262$ GeV$^2$ (upper points) and 
565 GeV$^2$ (lower points), plotted against $|t|$. 
The lines are $46.5\, e^{6.9t}$ and $33.5\, e^{5.5t}$.}
\vskip 9truemm
\centerline{\epsfxsize=0.40\hsize\epsfbox[70 60 395 285]{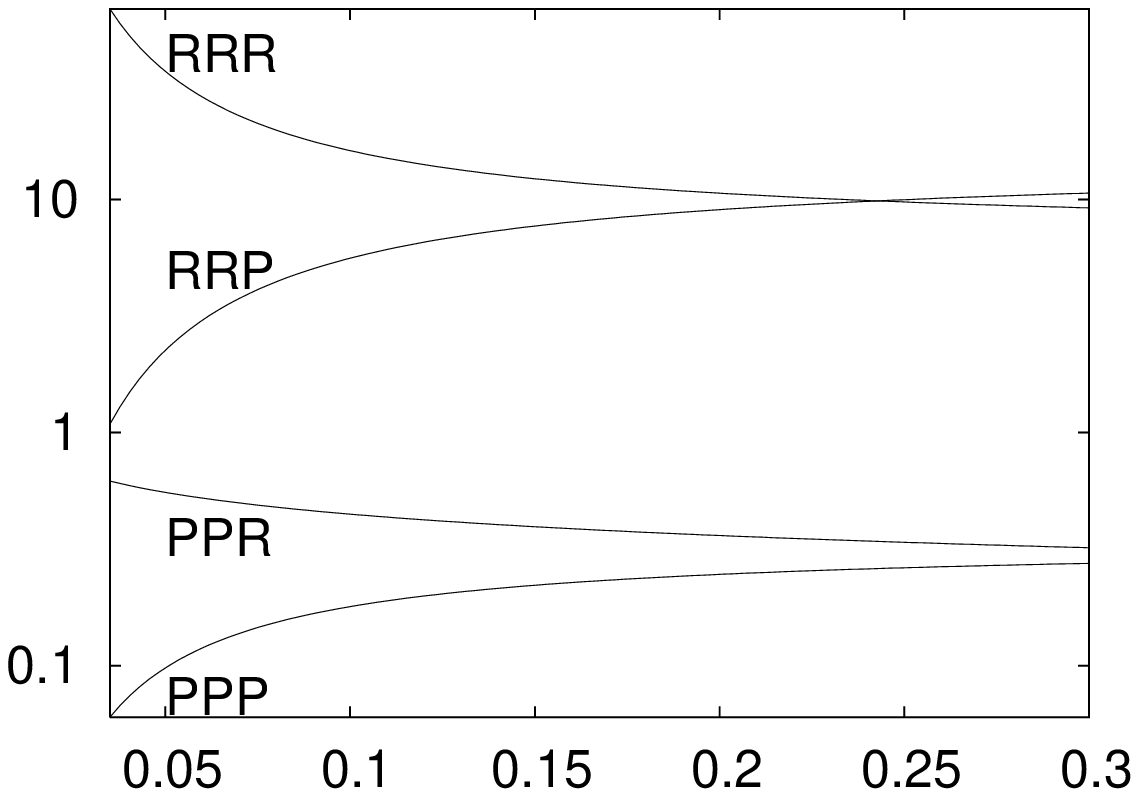}}
\vskip 2truemm
\centerline{\rfont Figure 3: the triple-reggeon vertices plotted against $|t|$}
\endinsert

\pageinsert
\line{\hfill
{\epsfxsize=0.40\hsize\epsfbox[50 50 395 290]{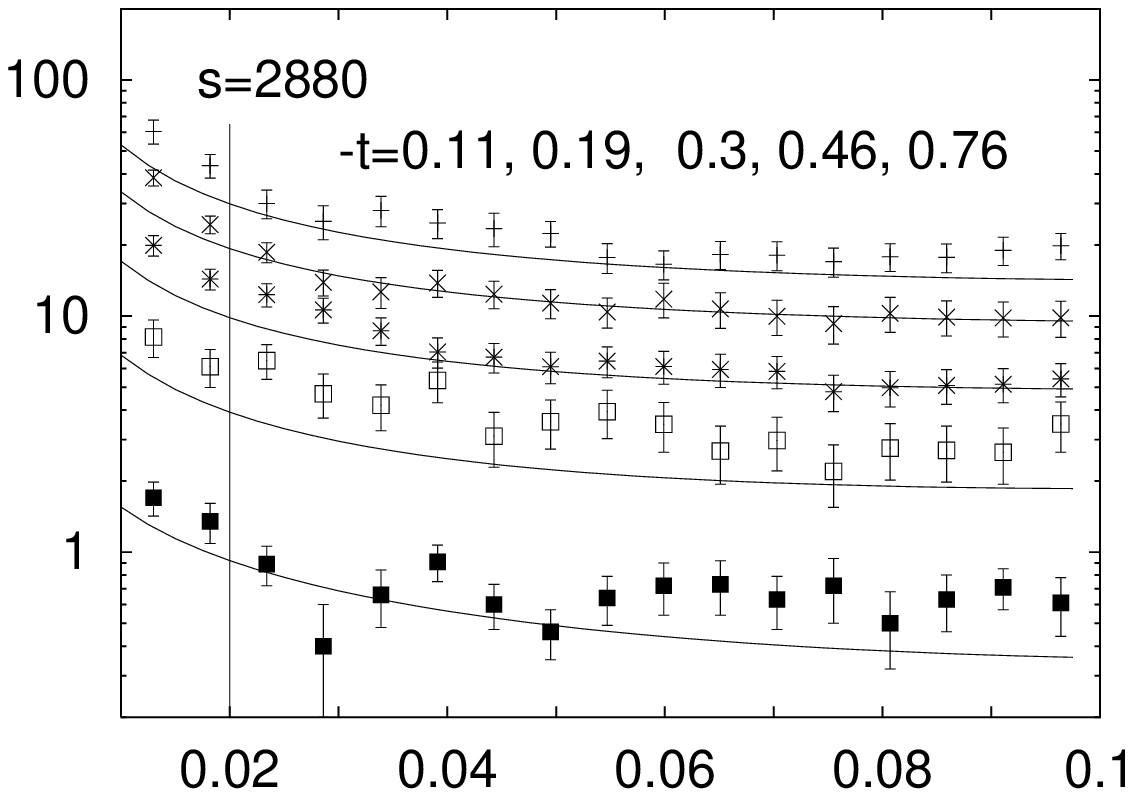}
\hfill
\epsfxsize=0.40\hsize\epsfbox[50 50 395 290]{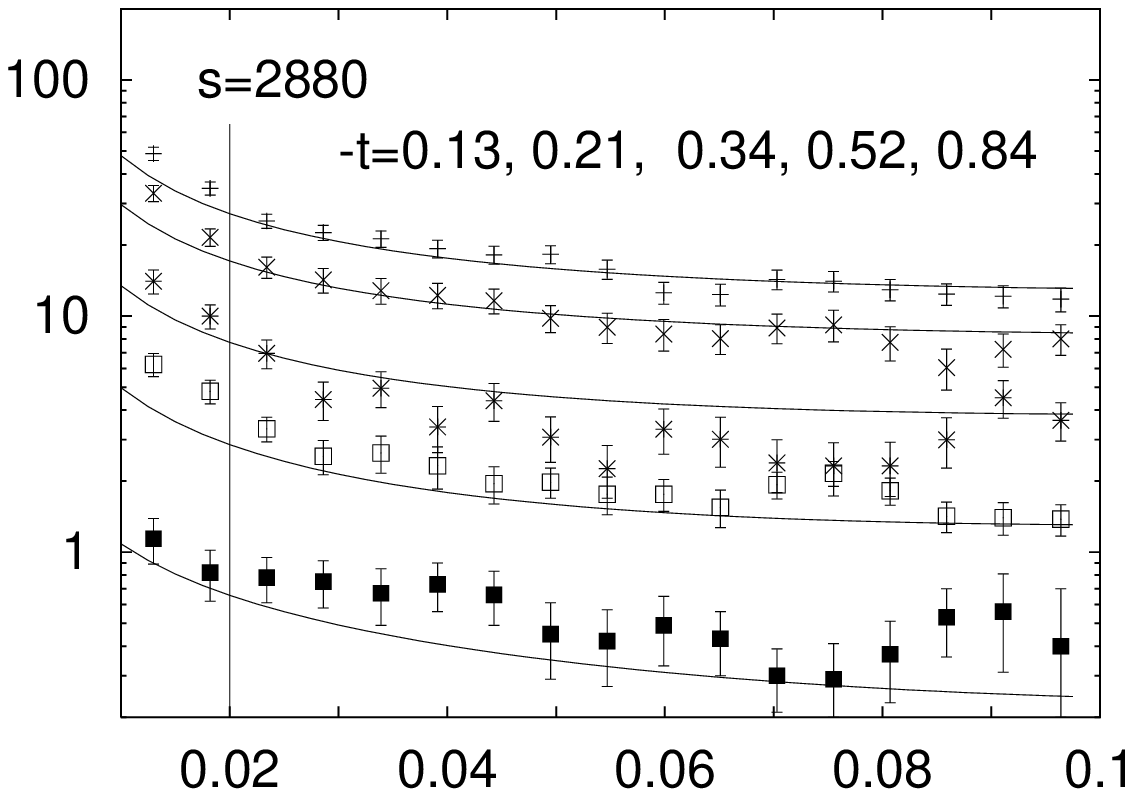}}\hfill}

\centerline{(a)$~~~~~~~~~~~~~~~~~~~~~~~~~~~~~~~~~~~~~~~~~~~~~~~~~~~~~~~~~~~~~~~~$(b)}
\vskip 2truemm
\line{\hfill
\epsfxsize=0.40\hsize\epsfbox[50 50 395 290]{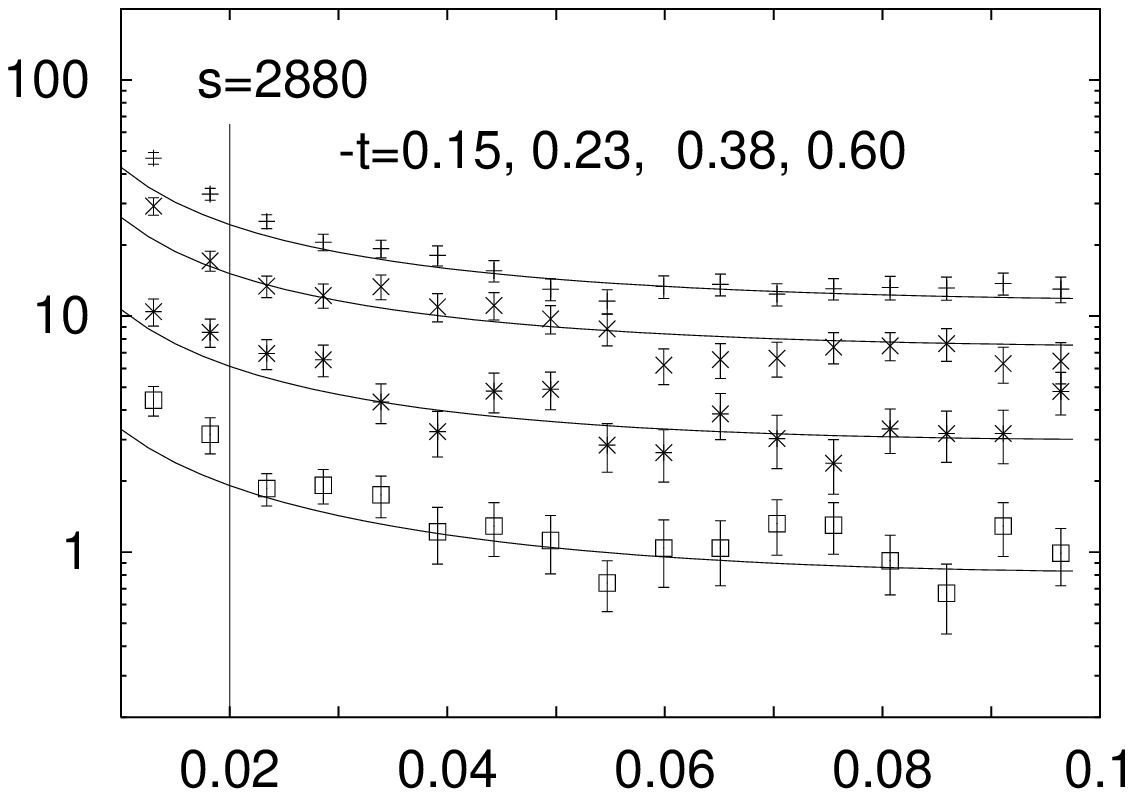}
\hfill
{\epsfxsize=0.40\hsize\epsfbox[50 50 395 290]{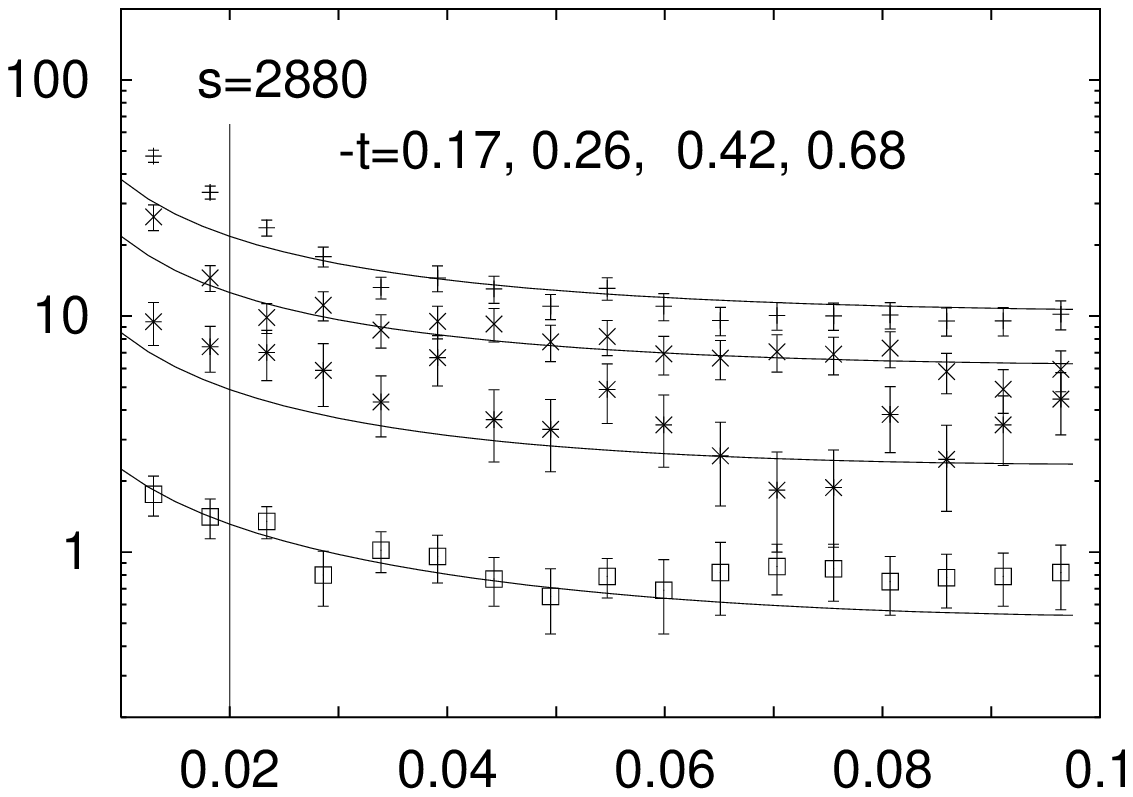}}\hfill}

\centerline{(c)$~~~~~~~~~~~~~~~~~~~~~~~~~~~~~~~~~~~~~~~~~~~~~~~~~~~~~~~~~~~~~~~~$(d)}
\vskip 2truemm
\line{\hfill
\epsfxsize=0.40\hsize\epsfbox[50 50 395 290]{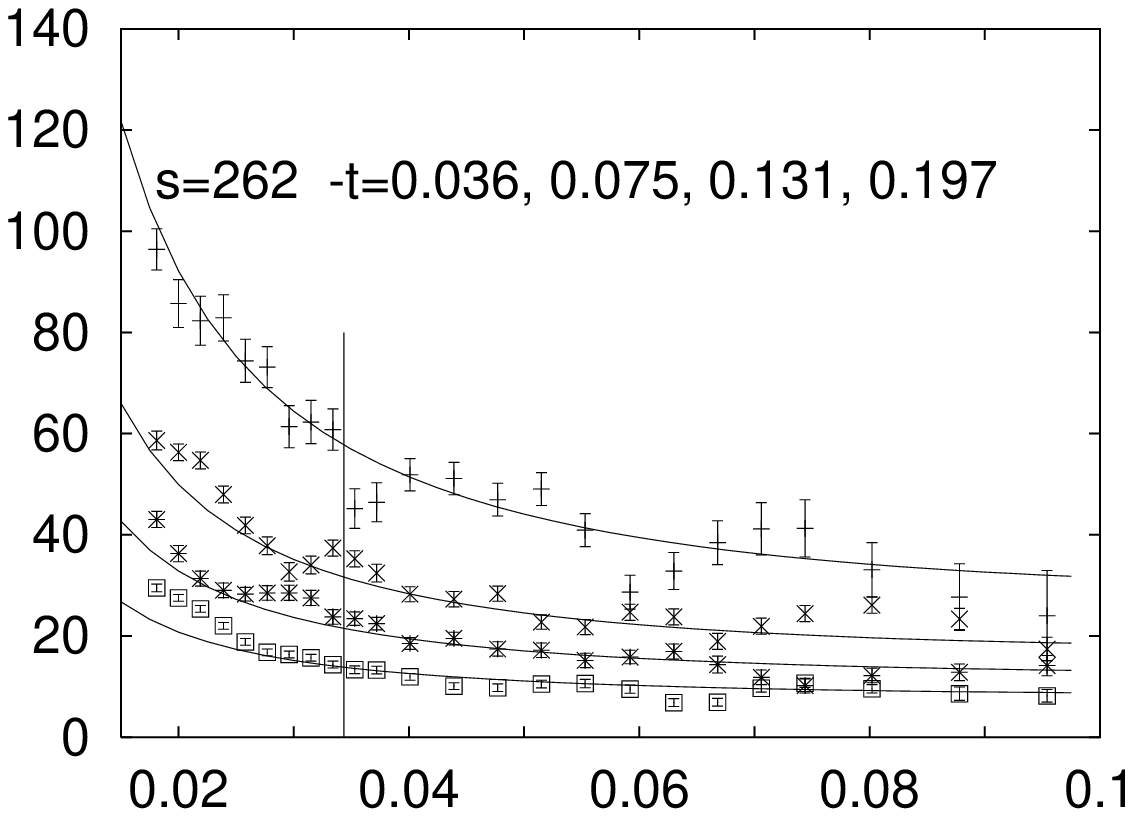}
\hfill
\epsfxsize=0.40\hsize\epsfbox[50 50 395 290]{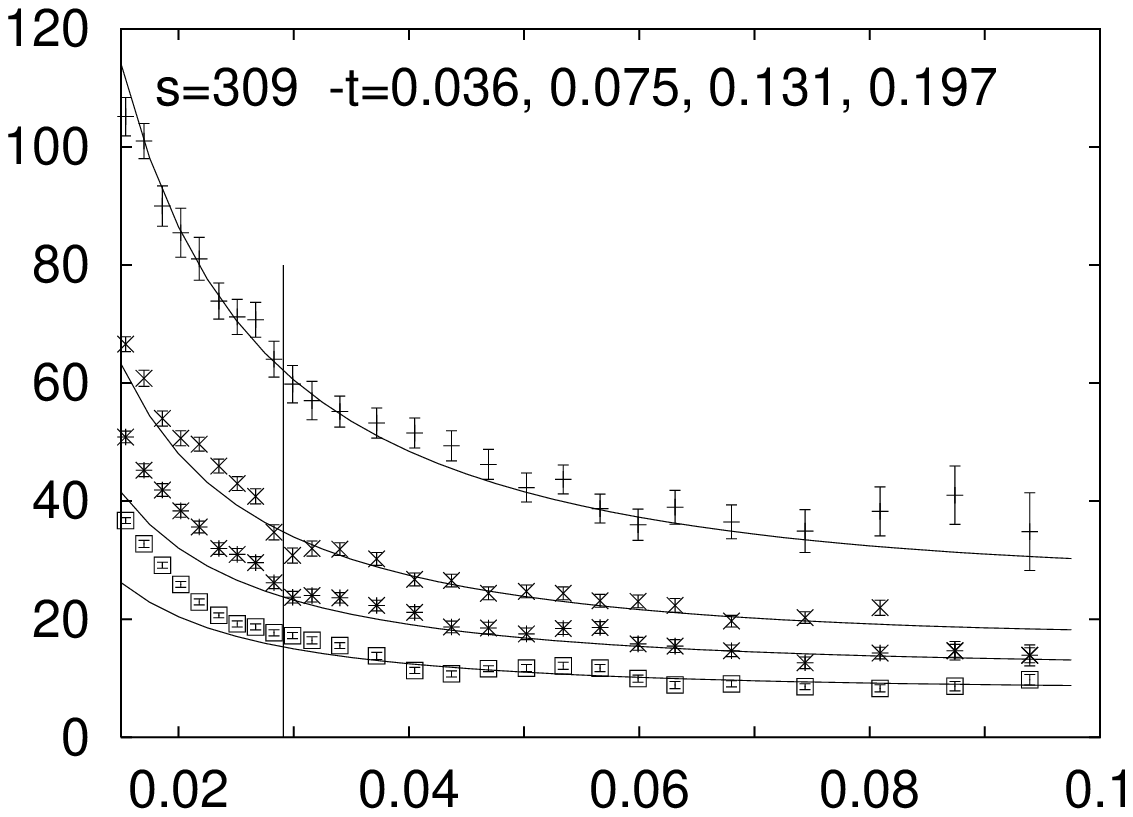}\hfill}

\centerline{(e)$~~~~~~~~~~~~~~~~~~~~~~~~~~~~~~~~~~~~~~~~~~~~~~~~~~~~~~~~~~~~~~~~$(f)}

\vskip 2truemm
\line{\hfill
\epsfxsize=0.40\hsize\epsfbox[50 50 395 290]{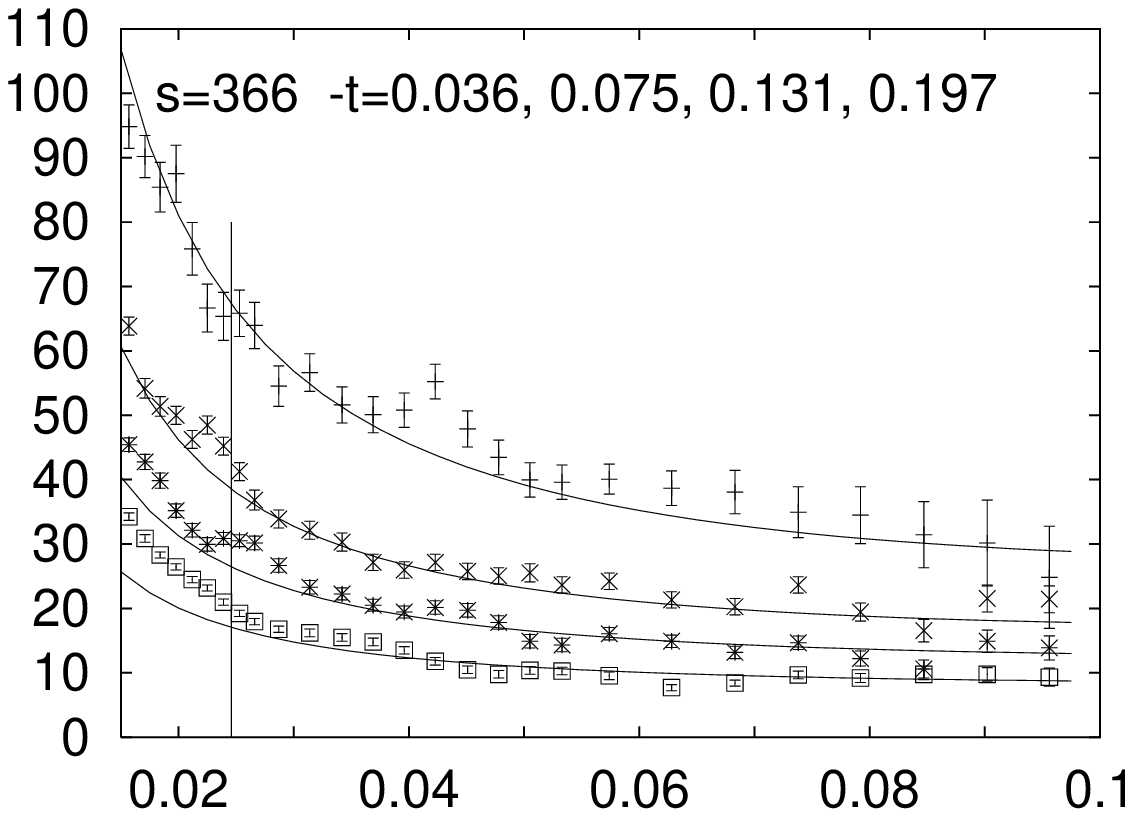}
\hfill
\epsfxsize=0.40\hsize\epsfbox[50 50 395 290]{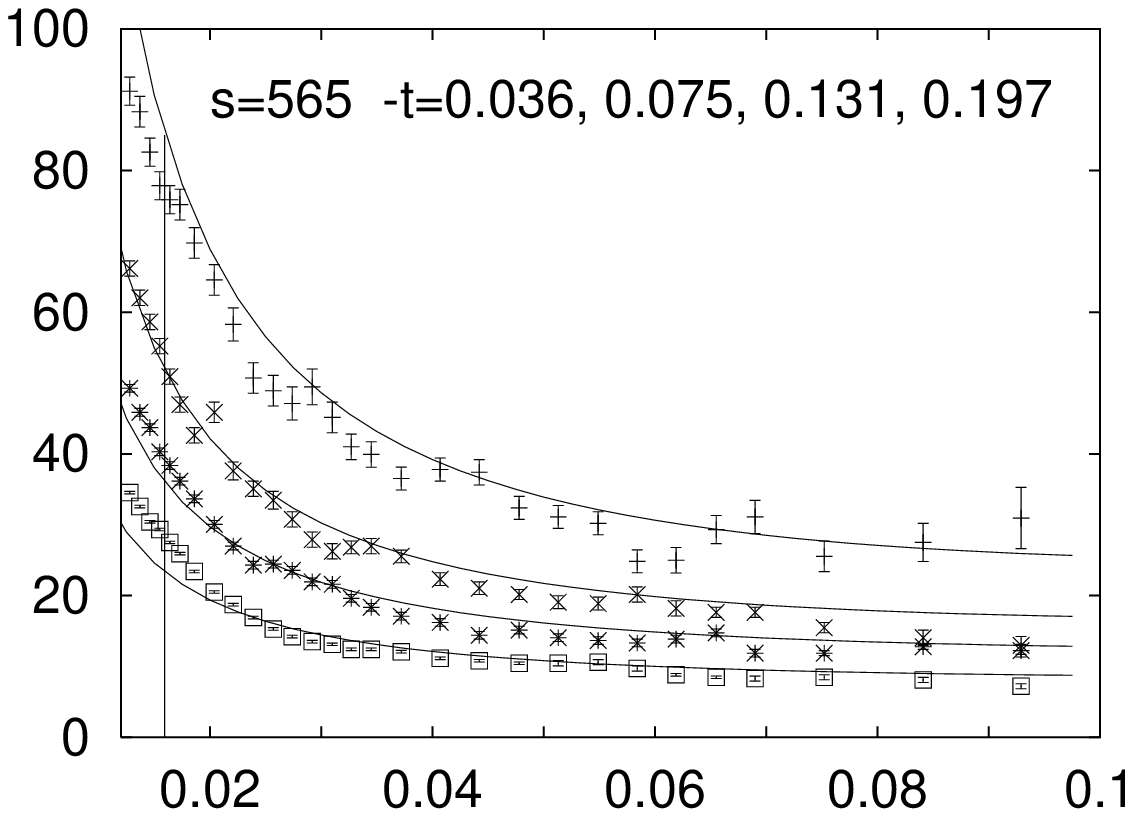}\hfill}

\centerline{(g)$~~~~~~~~~~~~~~~~~~~~~~~~~~~~~~~~~~~~~~~~~~~~~~~~~~~~~~~~~~~~~~~~$(h)}
\vskip 2truemm
{\rfont Figure 4: data\ref{\arm,\sch} for ${(1/\pi)d^2\sigma / dt\, d\xi}$ in
mb~GeV$^{-2}$ plotted against $\xi$
at various energies and values of $t$, compared with the fit. In each case,
only data points to the right of the vertical line are used in the fit.}
\endinsert

\topinsert
\vskip 2truemm
\line{\hfill
{\epsfxsize=0.37\hsize\epsfbox[50 50 395 290]{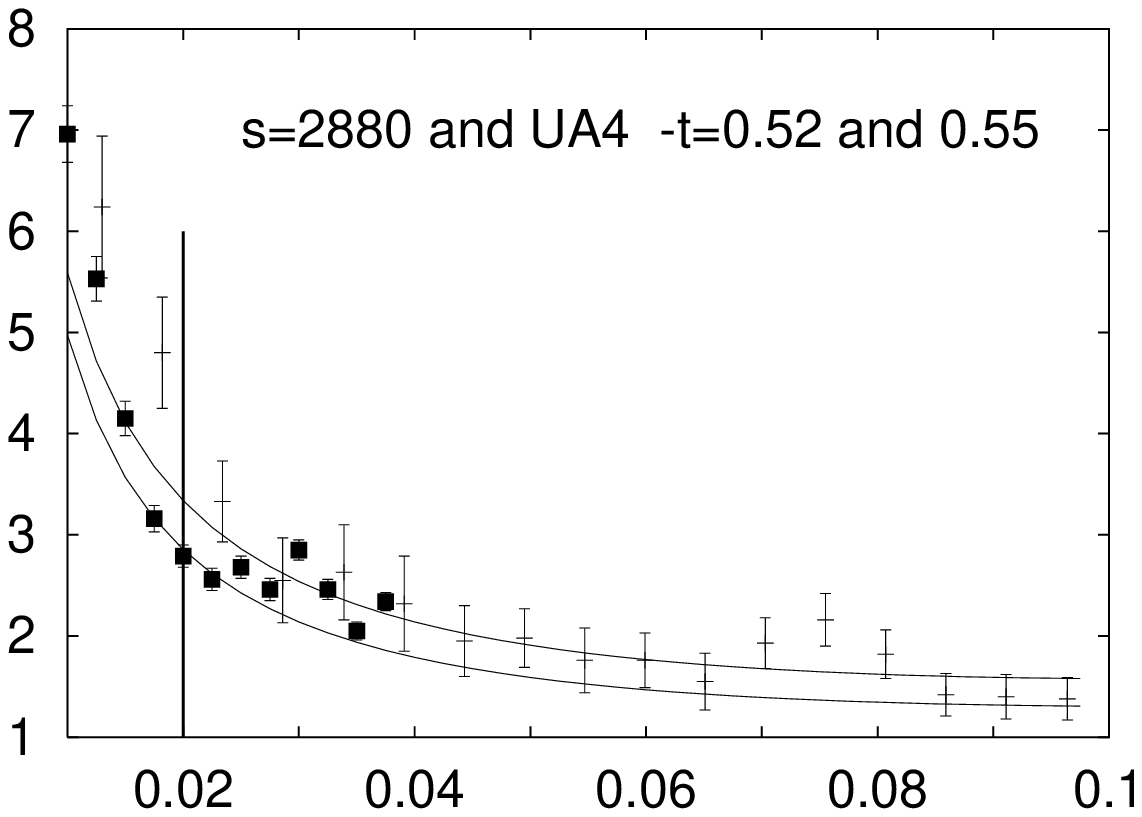}}
\hfill
{\epsfxsize=0.37\hsize\epsfbox[50 50 395 290]{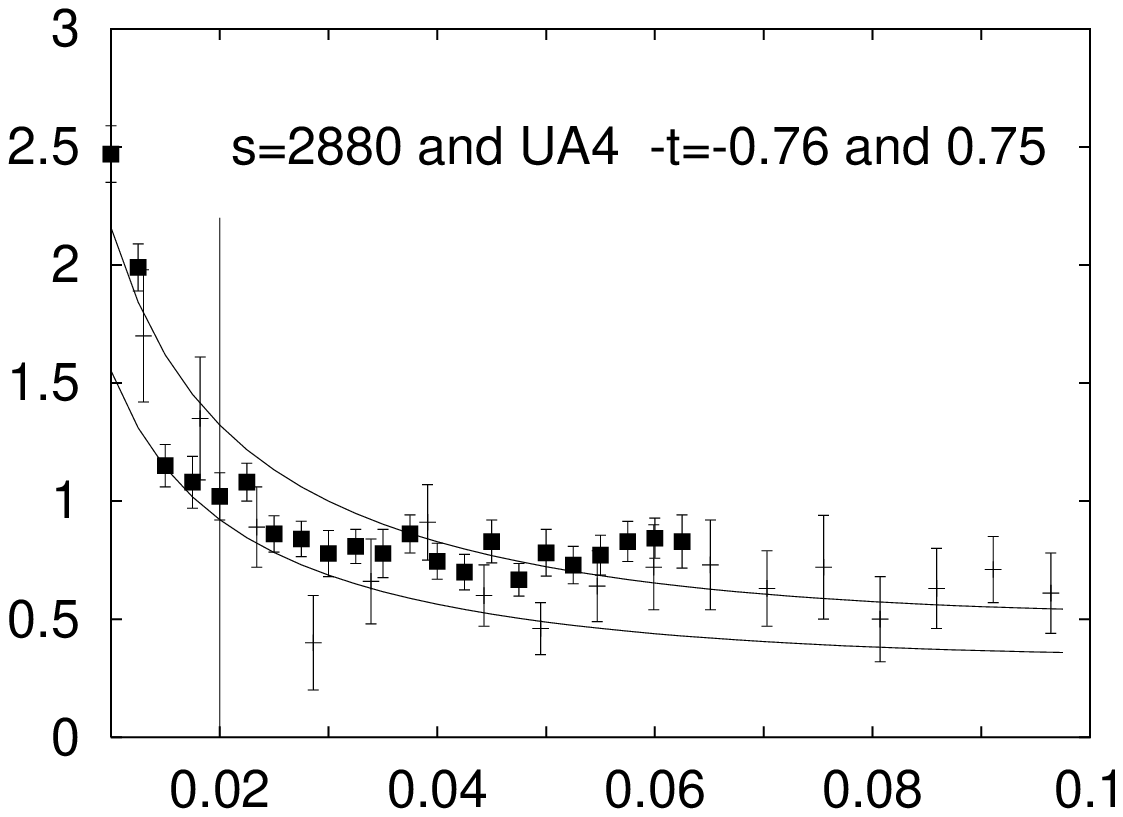}}\hfill
}

\centerline{(i)$~~~~~~~~~~~~~~~~~~~~~~~~~~~~~~~~~~~~~~~~~~~~~~~~~~~~~~~~~~~~~~~~$(j)}

\vskip 2truemm
{\rfont Figure 4 {\ritfont continued}: open points are from reference [{\albrow}], 
black points from UA4\ref{\uaf}. 
In each case the upper curve is at the higher energy.}
\endinsert

Our purpose is to show a reasonable fit to the data within conventional
triple-Regge theory, to demonstrate that it is not necessary to introduce
any new and unconventional ideas.
In order to keep things as simple as possible, we have just used the first
seven terms in (8), together with the pion terms (9). The latter contribute 
only at very small $t$ and are known, essentially without any free parameters:
$$
{d^2\sigma ^{\pi} \over dt\, d\xi}=D^{\pi}(t, \xi)~\sigma ^{\pi p}(M^2,t)
\eqno(10)
$$
with 
$$
D^{\pi}(t, \xi)={g^2_{\pi p}\over 16\pi^2}\,{t\over (t-m_{\pi}^2)^2}
(G(t))^2\xi^{1-2\alpha_{\pi}(t)}
$$$$
{g^2_{\pi p}\over 4\pi}=13.3~~~~~~~~~~\alpha_{\pi}(t)=\alpha'_{\pi}~(t-m^2_{\pi})
\eqno(11)
$$
For the trajectory slope we use $\alpha'_{\pi}=0.93$ GeV$^{-2}$. For the
form factor $G(t)$, which takes account of the proton wave function, we use
the Dirac form factor; this is surely not correct, but it will
not matter because ${d^2\sigma ^{\pi} / dt\, d\xi}$ is significant only
at very small $t$. For the same reason, we ignore the $t$ dependence of
$\sigma ^{\pi p}(M^2,t)$ and use its on-shell form\ref{\sigtot}
$$
\sigma ^{\pi p}(M^2,t)=13.63\, (M^2)^{0.08}+31.79\,(M^2)^{-0.45}
\eqno(12)
$$

We have used a linear $f_2$ trajectory\ref{\sigtot}
$$
\alpha_{R}(t)=1+\epsilon_{R}+\alpha'_{R}t
$$$$
\epsilon_{R}= -0.45~~~~~~~~~\alpha'_{R}= 0.93 \hbox{ GeV}^{-2}
\eqno(13)
$$
and an $f_2$ flux factor
$$
D^R(t,\xi)={9\beta_{R}^2\over 4\pi^2}\,(F_1(t))^2\,\xi^{1-2\alpha_R(t)}
\eqno(14)
$$
with $F_1(t)$ again the proton's Dirac form factor. We also use
$$
\sigma ^{\P p}(M^2,t)=V_{\P\P\P}(t)\,(M^2)^{0.08}+V_{\P\P R}(t)\,(M^2)^{-0.45}
$$$$
\sigma ^{R p}(M^2,t)=V_{RR\P}(t)\,(M^2)^{0.08}+V_{RRR}(t)\,(M^2)^{-0.45}
\eqno(15)
$$
where the $V_i(t)$ are the triple-reggeon vertices.

\topinsert
\line{\hfill
{\epsfxsize=0.37\hsize\epsfbox[50 50 395 290]{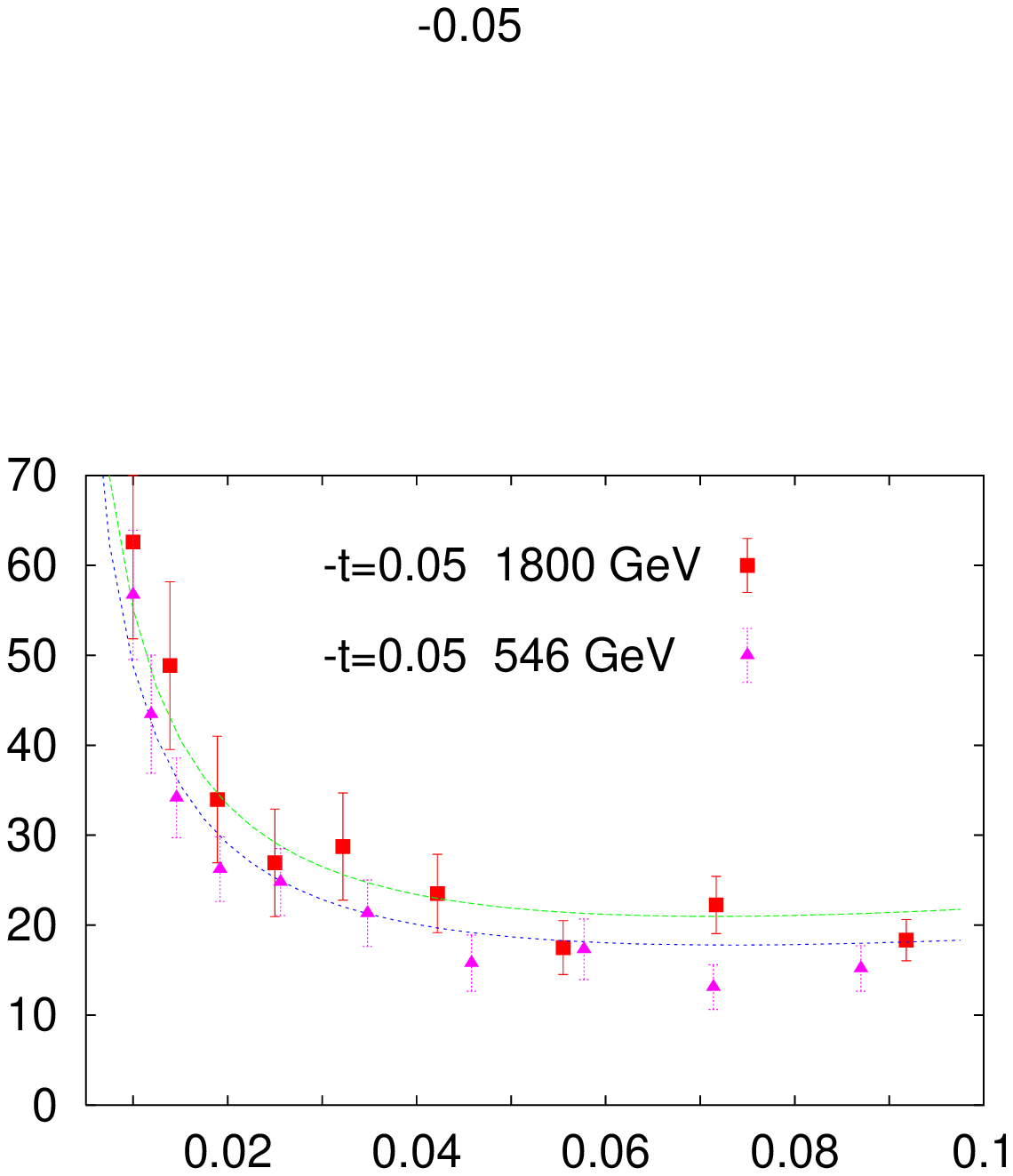}}\hfill
{\epsfxsize=0.37\hsize\epsfbox[50 50 395 290]{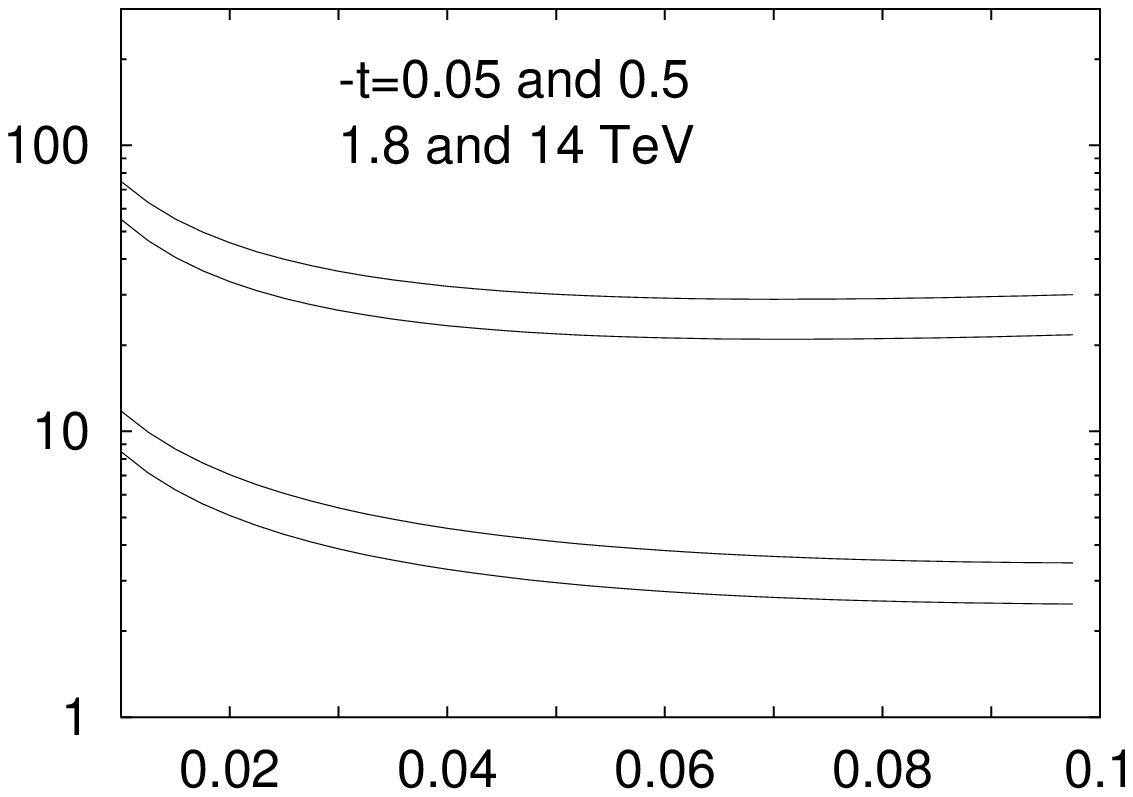}}\hfill
}

\centerline{(a)$~~~~~~~~~~~~~~~~~~~~~~~~~~~~~~~~~~~~~~~~~~~~~~~~~~~~~~~~~~~~~~~~$(b)}
\vskip 2truemm
{\rfont Figure 5: (a) reconstructed CDF data\ref{\goul}
for ${(1/\pi)d^2\sigma / dt\, d\xi}$ in
mb~GeV$^{-2}$ plotted against
$\xi$ with our fit;
(b) the rise of $(1/\pi)d^2\sigma / dt\, d\xi$ with $\sqrt s$ at 
$-t=0.05$ GeV$^2$
(upper pair of curves) and 0.5 GeV$^2$. In each case the upper curve is for the higher energy.
}
\endinsert

\topinsert
\line{\hfill\epsfxsize=0.45\hsize\epsfbox[80 60 390 290]{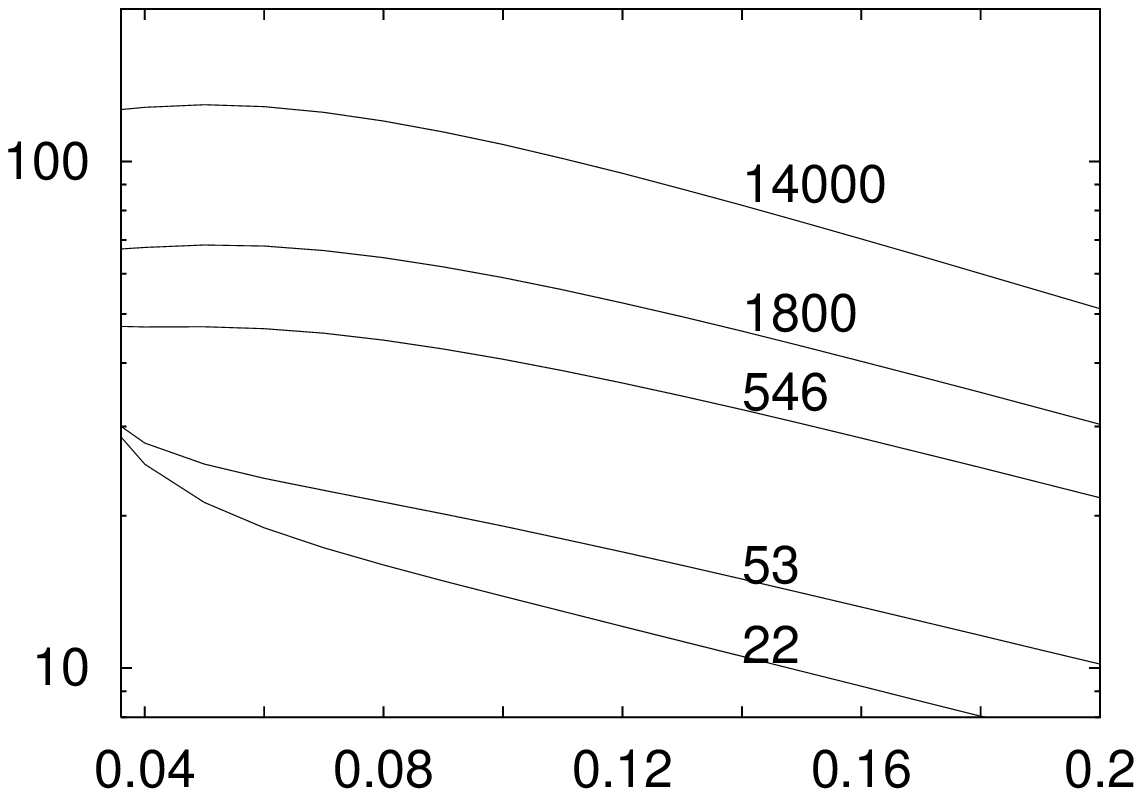}\hfill
\hfill\epsfxsize=0.45\hsize\epsfbox[75 50 380 285]{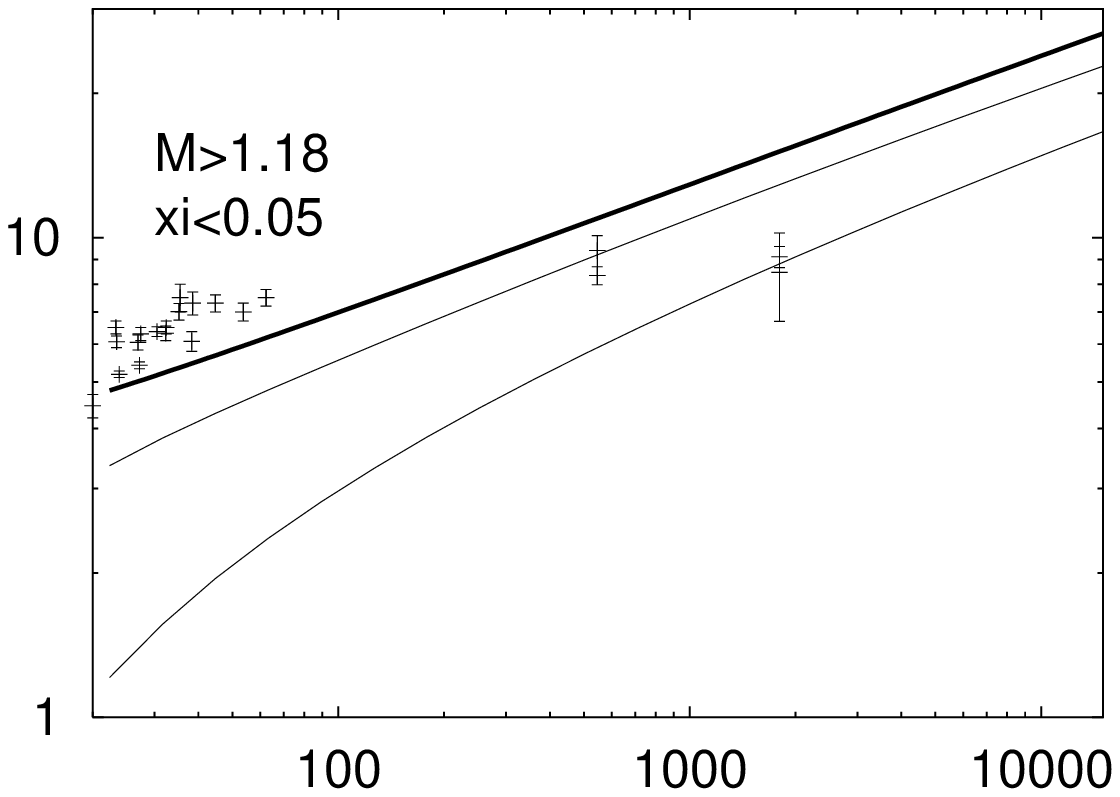}\hfill}

\centerline{(a)$~~~~~~~~~~~~~~~~~~~~~~~~~~~~~~~~~~~~~~~~~~~~~~~~~~~~~~~~~~~~~~~~$(b)}
\vskip 2truemm
{\rfont Figure 6: (a) $d\sigma /dt$ in
mb~GeV$^{-2}$ plotted against $|t|$ for various values
of $\sqrt s$; (b) $\sigma^{\hbox{{\fiverm DIFF}}}$ plotted against $\sqrt s$.
Both are for $\xi$ integrated up to 0.05 and down to $M^2=1.4$ GeV$^2$. In
(b)  the middle curve is for $-t$ integrated down to 0.036 GeV$^2$, the
lower curve is the contribution to it from ${\P\P\atop\P}$ and the
upper curve is an estimate for $-t$ integrated down to 0. The data are 
from the table in reference [{\goultwo}].}
\endinsert

It is well tested\defref\elastic{
A Donnachie and P V Landshoff, Nuclear Physics B267 (1986)  690
}
in $pp$ elastic scattering that it is the Dirac form
factor $F_1(t)$ that takes account of the proton wave function in the 
coupling of the pomeron to the proton. This is why we have included 
$F_1(t)$ in the pomeron flux factor $D(t,\xi)$ in (4). The coupling of
the reggeon to the proton is less well determined. Nevertheless, we are
free to include the same form factor in the reggeon flux factor
$D_R(t,\xi)$ in (14), because in the differential cross section
it is multiplied by triple-reggeon vertices. There
is no knowledge of how these four triple-reggeon vertices depend on
$t$, and the only way we can determine them is to fit their products
with the flux factors $D(t,\xi)$ or $D^R(t,\xi)$ to experiment. 
To this extent, what we choose for the $t$ dependence of the flux factors
is just a matter of convention; it is only when one introduces some
assumption about the $t$ dependence of the triple-reggeon vertices, or of
$\sigma ^{\P p}(M^2,t)$ and $\sigma ^{R p}(M^2,t)$, that the definitions
of the flux factors take on some independent significance.

Previous fits\ref{\robroy}\ref{\diffdis}\defref\hone{
H1 collaboration, C Adloff et al, Zeitschrift f{\"u}r Physik C74 (1997) 221
} have included large interference terms, and so we have included
such terms maximally. By this we mean
$$
{\P R\atop\P}+{R \P\atop\P}=2\sqrt{\Big ({\P\P\atop\P}\cdot{RR\atop\P}\Big )}
\,\cos(\half\pi(\alpha_{\P}(t)-\alpha_R(t))
$$$$
{\P R\atop R}+{R\P\atop R}=2\sqrt{\Big ({\P\P\atop R}\cdot{RR\atop R}\Big)}
\,\cos(\half\pi(\alpha_{\P}(t)-\alpha_R(t))
\eqno(16)
$$
The cosines arise from the Regge signature factors\ref{\book}.

We have experimented with various simple forms for the triple-reggeon vertices
and have arrived at
$$
V_i(t)=A_i\,\Big({t\over t+\mu^2_{i}}\Big )^{\lambda_{i}}
~~~~~~~~~~~~~i=\P\P\P\, , \P\P R\, , RR\P\, , RRR
\eqno(17)
$$
These are multiplied by the flux factors, which include
the couplings $\beta_{\P}^2$ or $\beta_R^2$.
We have varied the 12 parameters $\beta_i^2A_i, \,\mu^2_{i}$ and ${\lambda_{i}}$
to make our fit.
It has $\lambda_{RRR}$ quite large and negative, and $\lambda_{\P\P R}$ also
negative.
One should not conclude from this that the data require that the
corresponding vertices
diverge at $t=0$, because the smallest value of $t$ for which there is
data is $-t=0.036$ GeV$^2$ and it is only for $|t|$ larger than this that
our fit has any meaning. As may be seen in figure 3, our vertices all
behave sensibly. The reason that $V_{RRR}$  needs to vary fairly
rapidly at small $t$ may be seen in figure~2: at least at low energy, 
the data fall steeply with $|t|$ at very small $t$ and then flatten off.
It has been known for a long time\defref\kaidalov{
A B Kaidalov, Physics Reports 50 (1979) 157
}
that $V_{\P\P\P}(t)$ is small at small $t$.

In figures 4 and 5a we compare our fit with the data we have used, except that
we have not shown the $s=741$ GeV$^2$ data from reference [{\sch}] because they
are rather ragged. 
Note that our fit succeeds in reproducing the fact that the 
data at very small $t$ are higher at low energies than at the Tevatron.
Figure 5b shows our expectations for LHC energy; these predictions should be
regarded as qualitative because, as we have explained, there is not a sufficient
quantity of good lower-energy data to make a definitive fit.
 
Figure 6a shows $d\sigma /dt$. Although our fit used only data for
$M^2 > 9$ GeV$^2$, we have extrapolated it down to 1.4 GeV$^2$ and integrated
over $\xi$ down to this value and up to $\xi=0.05$. We have multiplied by
2 to account for both $pp\to pX$ and $pp\to Xp$ (or $p \bar p\to pX$ and 
$p\bar p\to X\bar p$). We show the plot only down as far as $-t=0.036$~GeV$^2$,
for the reasons we have explained.  We have explained also that it is
risky to extrapolate the simple fit to values of $M$ below $M_0=3$ GeV.

Figure 6b shows the integral of
$d\sigma / dt$. The middle curve integrates it down to 
$-t=0.036$~GeV$^2$ and the lower curve shows how much comes from the term
${\P\P\atop\P}$. We have extrapolated the curves in figure 6a down to
$t=0$ to get a rough estimate of how much we should add to the middle curve
to make it extend to $t=0$. This extrapolation is highly uncertain. If we
assume that between $t=0$ and $-t=0.054$ GeV$^2$ $d\sigma /dt$ may be well
approximated by a single exponential $e^{bt}$, so that
$$
\sigma(t_{\hbox{{\fiverm MIN}}}=0)=
(\sigma(t_{\hbox{{\fiverm MIN}}}=0.036))^3
/(\sigma(t_{\hbox{{\fiverm MIN}}}=0.054))^2
\eqno(18)
$$
this results in the upper curve. 
If instead we had used the 
parametrisation (17) with our large negative $\lambda_{RRR}$ it would have
led to a divergent cross section. 

The upper curve in figure 6b fits well to
$$
\sigma^{\hbox{{\fiverm DIFF}}}(s;M_{\hbox{\fiverm{MIN}}}^2=1.4\hbox{ GeV}^2)
=1.9\, s^{0.1425}
\eqno(19)
$$
in mb-GeV units. The data are from reference [{\goultwo}]. We must emphasise
that experiments all have limited acceptance in $t$ and $\xi$, so that
the values they quote for $\sigma^{\hbox{{\fiverm DIFF}}}(s;M_{\hbox{\fiverm{MIN}}})$ are more than a little model-dependent.
As we have just explained, our fit to the available data, which are for
$-t\ge 0.036$ GeV$^2$, can give very different total ``diffractive'' cross
sections when it is extrapolated to $t=0$ in different ways. Because we
do fit the available differential-cross-section data, the total 
diffractive cross section represented by the upper curve in figure 6b
is just as believable as the claimed data points; there is no trustworthy
extraction of the total diffractive cross section. This in turn leads to 
uncertainty in the results for $\sigma^{\hbox{\fiverm{TOTAL}}}$ when it is
extracted from collider experiments by the so-called luminosity-%
independent method\defref\cdftot{
CDF collaboration, F Abe et al, Physical Review D50 (1994) 5550
}.

Our fit (6)  to the total cross section 
predicts \hbox{$\sigma^{\hbox{{\fiverm TOTAL}}}=101.5$ mb} 
at $\sqrt s=14$ TeV,
while (19) gives $\sigma^{\hbox{{\fiverm DIFF}}}=29$ mb,
so there is no need to introduce significant shadowing up to this energy.

As is seen in figure 6b, even at Tevatron energy the term ${\P \P\atop\P}$
contributes less than 70\% of the total ``diffractive'' cross section. 
At very large energies this fraction rises to nearly 80\%. Accurate 
measurement of $d^2\sigma/dtd\xi$ at RHIC would be of great value in
clarifying further the details of the diffractive dissociation mechanism. 
\vskip 10truemm
{\sl This research was supported in part by PPARC}
\vskip 10truemm
\parskip =2pt
\medskip\immediate\closeout\rfile\writestoppt
\baselineskip=14pt{{\bf References}}\bigskip{\frenchspacing%
\parindent=20pt\escapechar=` \input refs.tmp\bigskip}\nonfrenchspacing
\bye